\UseRawInputEncoding
\documentclass{article}

\usepackage{arxiv}

\usepackage[utf8]{inputenc} % allow utf-8 input
\usepackage[T1]{fontenc}    % use 8-bit T1 fonts
\usepackage[hyphens]{url}
\usepackage{hyperref}
\hypersetup{colorlinks,allcolors=black}
\hypersetup{breaklinks=true}
\usepackage{cite}
\usepackage{booktabs}       % professional-quality tables
\usepackage{amsfonts}       % blackboard math symbols
\usepackage{nicefrac}       % compact symbols for 1/2, etc.
\usepackage{microtype}      % microtypography
\usepackage{lipsum}
\usepackage{graphics}
\usepackage{amssymb}
\usepackage[overload]{empheq}
\usepackage[colorinlistoftodos]{todonotes}
\usepackage[square,numbers]{natbib}
\usepackage{comment}
\usepackage{tabularx}
\usepackage[nobottomtitles*]{titlesec}
\usepackage{subfig}

\title{How mass surveillance can crowd out installations of COVID-19 contact tracing apps}

\author{
  Eran Toch\thanks{http://toch.tau.ac.il} \\
  Tel Aviv University\\
  \texttt{erant@tauex.tau.ac.il} \\
   \And
 Oshrat Ayalon\thanks{https://www.oshratayalon.com/} \\
  Max Planck Institute for Software Systems\\
  \texttt{oayalon@mpi-sws.org} \\
}

\begin{document}
\maketitle

\begin{abstract}
During the COVID-19 pandemic, many countries have developed and deployed contact tracing technologies to curb the spread of the disease by locating and isolating people who have been in contact with coronavirus carriers. Subsequently, understanding why people install and use contact tracing apps is becoming central to their effectiveness and impact. This paper analyzes situations where centralized mass surveillance technologies are deployed simultaneously with a voluntary contact tracing mobile app. We use this parallel deployment as a natural experiment that tests how attitudes toward mass deployments affect people's installation of the contact tracing app. Based on a representative survey of Israelis (n=519), our findings show that positive attitudes toward mass surveillance were related to a reduced likelihood of installing contact tracing apps and an increased likelihood of uninstalling them. These results also hold when controlling for privacy concerns about the contact tracing app, attitudes toward the app, trust in authorities, and demographic properties. Similar reasoning may also be relevant for crowding out voluntary participation in data collection systems.
\end{abstract}

\section{Introduction}
The COVID-19 pandemic is an unprecedented global crisis that poses a severe threat to the health and well-being of every person on the planet. In their efforts to control the epidemic, many countries have turned to contact tracing technologies as one of the main tools to curb the spread of the virus. For decades, rapid contact tracing has been used as an effective public health response in the face of infectious disease outbreaks. Successful contact tracing is based on identifying people who have come in contact with infected people and then quarantining them to interrupt further transmission of the epidemic \cite{fraser2004factors}. COVID-19 presents a problem for traditional contact tracing because many transmissions happen early in the infection cycle, before the onset of symptoms and before test results are received. Therefore, many countries have begun to develop technologies that capture "proximity events", in which two mobile phones are close enough for sufficient time for the risk of infection to be inferred.

In the months that followed the pandemic's outbreak, many countries, including South Korea, Singapore, and Israel, introduced contact tracing mobile applications, which detect proximity through Bluetooth or colocation \cite{ahmed2020survey}. Many other countries soon converged to decentralized architectures that do not allow third parties to learn which places users have visited or which people they have met \cite{canetti2020private}. After several suggestions for privacy-enhancing blueprints for contact tracing technologies \cite{canetti2020private}, Apple and Google announced a Bluetooth-based protocol to be embedded in both iOS and Android operating systems. Given the desire for population-wide dissemination, a small number of countries have deployed contact tracing based on involuntary mass surveillance \cite{jalabneh2020use,Altshuler2020Digital}). Unlike voluntary apps, which require individuals to actively install a contact tracing app, involuntary systems rely on mass surveillance to infer contacts without individual consent. Nonvoluntary solutions often rely on centralized acquisition of cellphone location data from cellular providers. For example, South Korea \cite{shaw2020governance} and Israel \cite{amit2020mass} used cellular traces from mobile carriers to track contacts.

The effectiveness of contact tracing technologies depends heavily on people's behavior, particularly on the proportion of people who install and use the technology. The greater the number of people who install the app, the more health authorities can identify potential contacts and the easier it is to effectively control the epidemic \cite{ferretti2020quantifying,cencetti2020digital}. Significantly, installation rates among the general population can indirectly benefit more vulnerable people, such as the elderly \cite{lopez2021anatomy}. Several empirical studies show that CTTs can be effective in curbing the spread of COVID-19 \cite{kendall2020epidemiological,fetzer2020does,wymant2021epidemiological}. Based on the actual installation rates of apps around the world, ranging between 15\% and 40\% \cite{Altshuler2020Digital}, the attention of many public health scholars has turned to the reasons that acceptance rates are not higher.

Understanding why people install contact tracing apps has led researchers to study people's attitudes in the U.S. \cite{zhang2020americans,li2021makes,kaptchuk2020good,seberger2021us}, South Korea \cite{kim2021covid}, Japan \cite{machida2021survey}, France \cite{guillon2020attitudes}, Germany \cite{meier2021investigating}, the Netherlands \cite{jonker2020covid}, Ireland \cite{fox2021exploring}, and Switzerland \cite{von2021drivers}, as well as through international comparison \cite{altmann2020acceptability,chan2021privacy,bonner2020lacking,utz2021apps,chan2021privacy}. People's attitudes toward contact tracing technologies are shaped by their perceptions of the health benefits they offer, to them individually or to the people around them (e.g., \cite{altmann2020acceptability,chan2021privacy,fox2021exploring}). Surveys also found a link between trust in health governance and acceptance of CCTs \cite{guillon2020attitudes}. Similarly, many surveys have found a connection between higher privacy concerns and lower acceptance of CCTs, ranging between strong correlations \cite{zhang2020americans,kaptchuk2020good,chan2021privacy,altmann2020acceptability} and weaker links \cite{li2021makes,fox2021exploring}.

%This is particularly important given the fact that cellular tracking can be very inaccurate, in comparison to mobile apps that can use Bluetooth and more precise locations for contact tracing [I will add a citation]. This behavior may be influenced by peoples perceptions of the technologies utility [I will add related-works] or by their perceptions of the potential privacy threats [I will add related-works] that may originate from personal information collection. Specifically, we ask whether deploying an involuntary system affect the deployment of voluntary system.

%The argument against non-voluntary technologies mostly concentrate on the substantial negative impact on citizens privacy and on the involvements of survillence  forces in what is, essentially, a health challenge \cite{eck2020state,marciano2021israel}.

Many interventions against the COVID-19 pandemic share the fundamental dilemma of voluntary action, which relies on people's individual decisions versus involuntary solutions that do not require active participation. While compulsory methods may be more effective, there is a basis for suspecting that they may crowd out motivations for voluntary participation. Multiple behavioral experiments show that enforcement can reduce intrinsic motivation, a phenomenon termed "motivational crowding out" \cite{bowles2012economic}. As involuntary mandates increase, voluntary contributions are increasingly crowded out, even when there is a personal benefit to participation. For example, field experiments have documented adverse spillover effects of monitoring workers' productivity \cite{belot2016spillover}. In the context of the COVID-19 pandemic, studies have documented the impact of crowding out on social distance measures \cite{yan2021measuring} and on people's acceptance of different types of countermeasures \cite{schmelz2021enforcement}. The link between mass surveillance and passivity is particularly disturbing for the public's engagement in taking the voluntary steps necessary to combat the COVID-19 pandemic.

This paper explores the crowding-out spillover effects that involuntary mass surveillance systems may have on voluntary contact tracing app installations. To operationalize this hypothesis in the context of the rapidly evolving COVID-19 pandemic, we empirically examine whether attitudes toward the involuntary system are associated with the likelihood of installing the official Israeli contact tracing apps. We use Israel's deployment of two contact tracing technologies in the spring of 2020 as a natural experiment for assessing the adoption of existing contact tracing app and attitudes toward variants of contact tracing technologies. The first is \emph{HaMagen} ("the Shield", in Hebrew), a privacy-preserving contact tracing mobile app that the Ministry of Health developed. The second is a mass surveillance technology-based cellular tracking technology operated by Israel's General Security Services (GSS), dubbed "The Tool." The tool uses a mixture of GPS locations transmitted through cellular protocols and cellular antenna triangulation to track the location of the whole population \cite{nytimes2020track}. Regularly, The Tool is used according to authorization from a court order \cite{security2002law}, but in March 2020, the Israeli government authorized the use of this technology for contact tracing \cite{nytimes2020track}. A detailed description of the technologies, as well as a timeline of their deployments, is available at Appendix A.

Given the theory on crowding out, we ask whether the existence of mass and involuntary surveillance can interfere with the deployment of voluntary technologies and reduce the installations of contact tracing apps. We operationalize this general research inquiry to two specific questions: What is the effect of attitudes toward mass surveillance on people's (1) installation of the contact tracing app and the (2) uninstallation of the app?

\section{The Survey and Results}
To study the possibility that involuntary contact tracing may crowd out voluntary contact tracing, we conducted a representative online survey of Israelis between May 4 and May 7, 2020. The survey was conducted 49 days after the mass surveillance technology was deployed in Israel and 43 days after the contact tracing app was deployed (see complete timeline of deployments in Appendix A). Participants were presented with questions about the contact tracing app and mass surveillance technology. For each technology, we asked the participants to read a short excerpt from a daily newspaper that describes the technology and answer several questions about their attitudes toward it. We have randomized the order of the two technologies. Detailed information about the survey is available at Appendix B and the breakdown of the participants' answers is available at Appendix C.

The final study sample includes 519 respondents. Our unweighted sample was nearly representative of the Israel population concerning gender, age, religion/ethnicity, religious level, and education. As shown in Table \ref{tab:demographics}, the mean age was 38 years (with a standard deviation of 13.80). A total of 53\% of the participants were women, 46\% were men, and the rest had chosen the "other" category. Religion or ethnic background was as follows: Jews (82\%), Arab Muslims (10.5\%), Arab Christians (2\%), and Druze (4\%). These results are in line with representative surveys carried out in Israel \cite{cooperman2016israel}. Other demographic variables that were recorded included religious level, education, and income.

A total of 166 out of our 519 participants had installed the app (32\%). This proportion is somewhat higher 
than the ratio of smartphone owners who had installed the app in Israel, which was 2.6 million people at November 2020 out of approximately 6.2 million smartphone owners \cite{altshuler2020contact}. 45 participants had reported uninstalling it (9\%), which is similar to the proportion of uninstallations.

\subsection{Contact Tracing App Installation}

\begin{figure}[ht]
\begin{minipage}{.5\linewidth}
\centering
\subfloat[]{\label{fig:att_margin}\includegraphics[scale=.5]{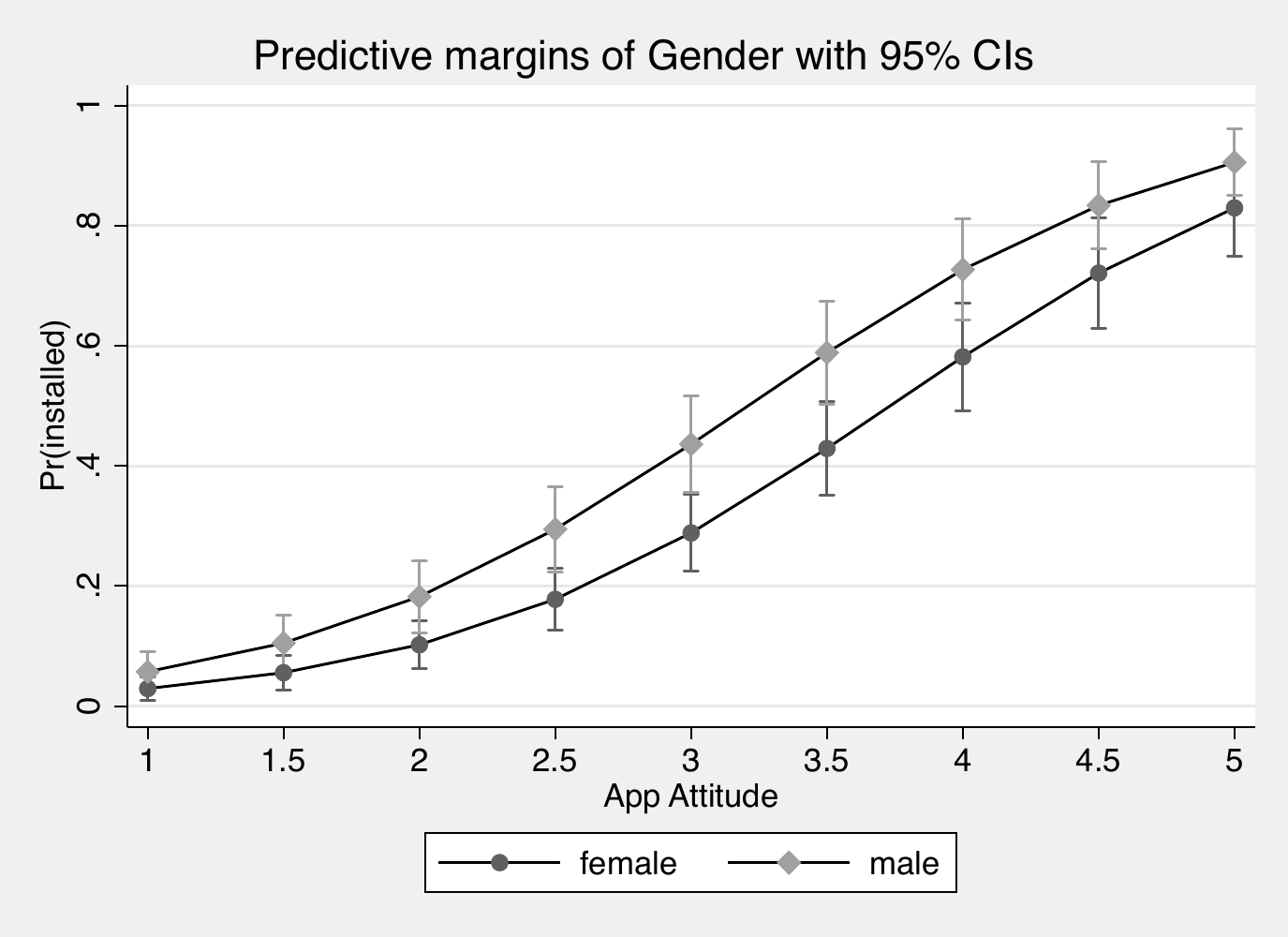}}
\end{minipage}%
\begin{minipage}{.5\linewidth}
\centering
\subfloat[]{\label{fig:cel_margin}\includegraphics[scale=.5]{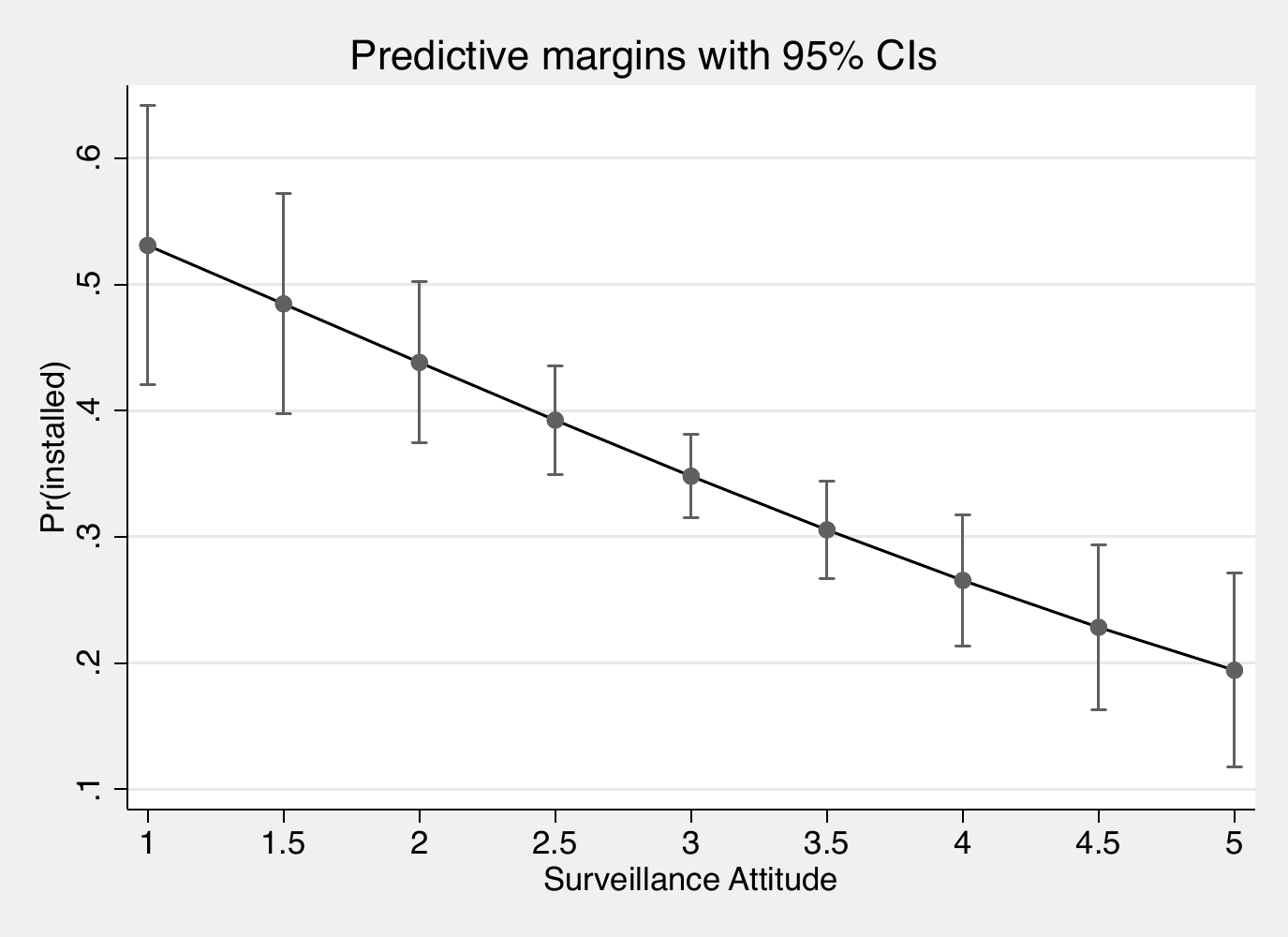}}
\end{minipage}\par\medskip
\centering
\subfloat[]{\label{fig:pri_margin}\includegraphics[scale=.5]{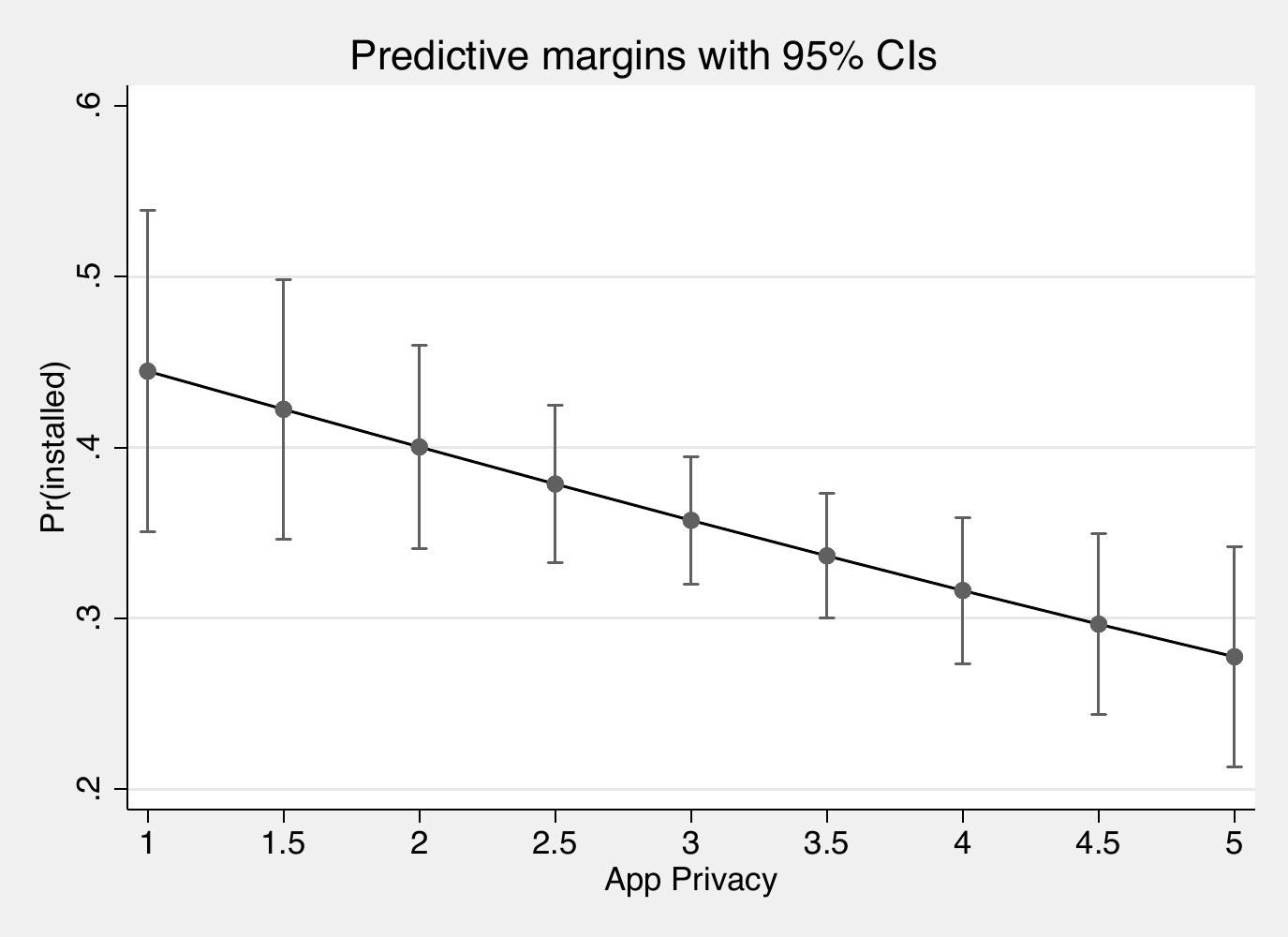}}
\caption{Relationship between installation of the app and amount of attitudes toward app (a), attitudes toward mass surveillance (b), and privacy concerns (c). Marginal probabilities are shown with 95\% CI.}
\label{fig:install_margin}
\end{figure}

To analyze the factors contributing to installing the application we ignored users who had uninstalled the app. We fitted a logistic regression model to the installation variable. The model fitted the data with a log-likelihood of -215.39 and a pseudo $R^2$ of 0.291. The model correctly classifies 78.56\% of the data points. The full model is presented in Table \ref{tab:install}. Figure \ref{fig:install_margin} provides an overview of the strongest factors associated with installation. We control for differences in observed heterogeneity by including gender, age, gender, education, religion, level of religious observance, and income effects.

The likelihood of installing the app is strongly correlated with the perceived attitude about the app (OR: 3.923, 95\% CI 2.918-5.274), as shown in Figure \ref{fig:install_margin} (a). Other measures of the app's utility were highly associated with this variable and were not significant with the model. We found that attitudes toward surveillance were the strongest and most significant negative factor associated with the likelihood of installing the app (OR: 0.547, 95\% CI 0.372--0.803). For every positive increase in attitude toward the utility of the mass surveillance system scale, there is a decrease of 45\% in the likelihood of installing the app (as seen in Figure \ref{fig:install_margin} (b)). This significant and negative correlation confirms our central hypothesis and points to an interaction between mass surveillance and contact tracing app installation. Privacy concerns about the contact tracing app are also negatively correlated with the likelihood of installation (OR: 0.758, 95\% CI 0.605-0.95). We did not find other associations to be significant. Specifically, concerns about the pandemic were not found to be significantly associated with installation or trust in leaders.

In terms of control variables, some of the demographic factors were related to app attitude or installation. We found gender to be the most robust significant variable with an effect on installations, with men approximately 100\% more likely to install the app (OR: 2.04, 95\% CI 1.215 -- 3.429). Younger people are more willing to install the app, but the relationship is generally weak and concentrated on the 18-20 year old demographic. People with graduate degrees are approximately 100\% times more likely to install the app (OR: 2.034, 95\% CI 0.984 -- 4.20), but the large confidence intervals point to a weak effect.

\subsection{Contact Tracing App Uninstallation}

\begin{figure}
\begin{minipage}{.5\linewidth}
\centering
\subfloat[]{\label{fig:att_margin_un}\includegraphics[scale=.5]{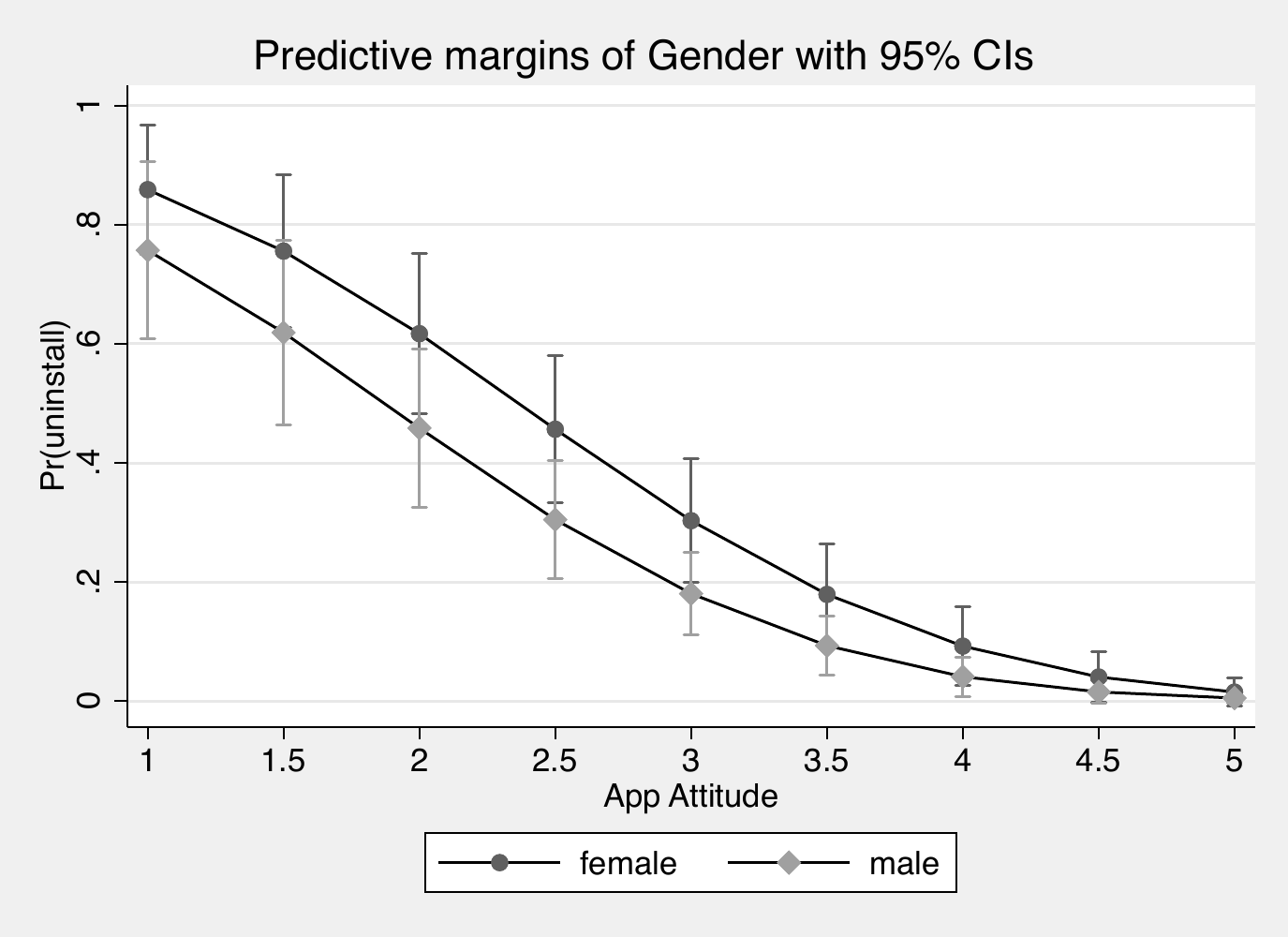}}
\end{minipage}%
\begin{minipage}{.5\linewidth}
\centering
\subfloat[]{\label{fig:cel_margin_un}\includegraphics[scale=.5]{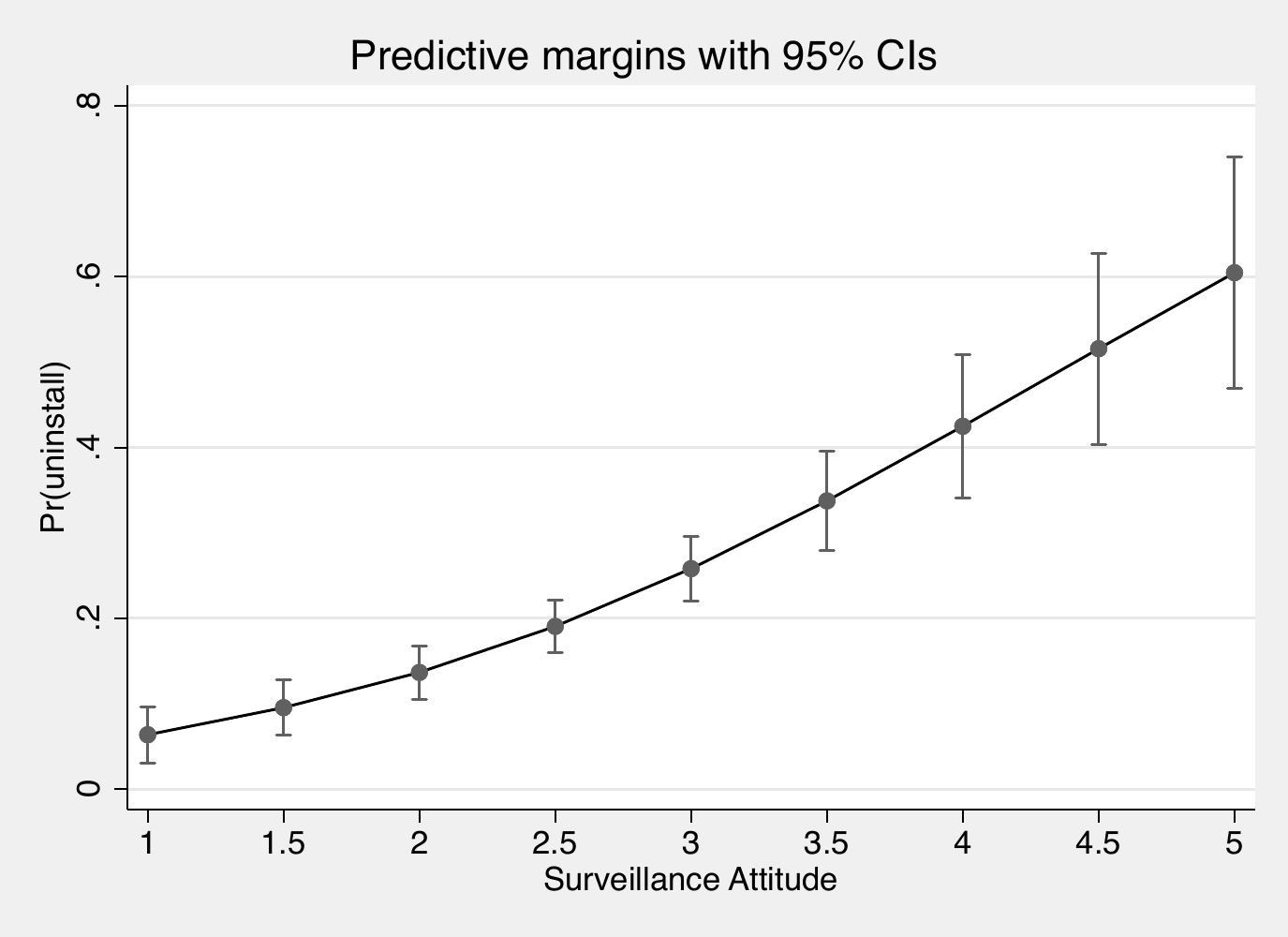}}
\end{minipage}\par\medskip
\centering
\subfloat[]{\label{fig:bat_margin}\includegraphics[scale=.5]{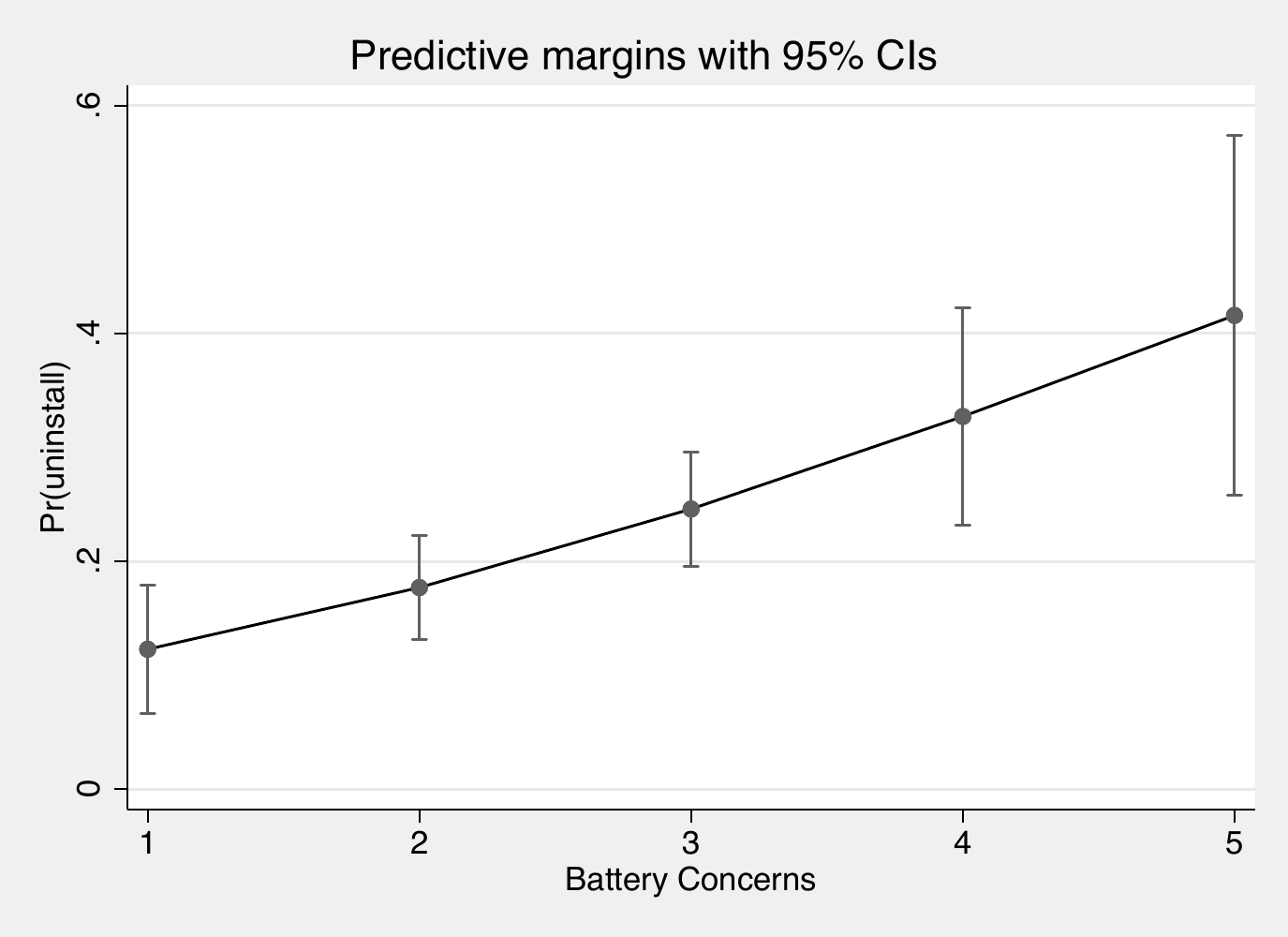}}

\caption{Relationship between uninstalling the contact tracing app and attitudes toward the app (a), attitudes toward mass surveillance (b), and battery concerns (c). Marginal probabilities are shown with 95\% CI.}
\label{fig:uninstall_margin}
\end{figure}

In our second research question, we look at the factors related to uninstalling the contact tracing app. We fitted a model for a dataset (n = 211) that contained only the people who either installed (166 people) or uninstalled the application (45 people). This proportion fits the proportion of people who had uninstalled it HaMagen at the time of the study \cite{altshuler2020contact}. Appendix D presents the full results of the logistic regression, and Figure \ref{fig:uninstall_margin} displays the marginal contribution of selected factors to the probability of uninstallation. The model fitted the data with a log-likelihood of -74.44 and a pseudo $R^2$ of 0.507. The model correctly classifies 89.32\% of the data points.

The most substantial factor associated with uninstalling the contact tracing app is attitude toward surveillance (OR: 8.57, 95\% CI 2.837-25.908). While the spread of this effect is relatively wide, on average, a user is approximately two times more likely to uninstall the app with every increase in the utility of the mass surveillance system, as seen in Figure \ref{fig:uninstall_margin} (b). The second most powerful reason for uninstalling the app is related to concerns about battery consumption (OR: 2.23, 95\% CI 1.28-3.88); users are approximately 120\% more likely to uninstall the app for every increase in these concerns. Other variables, such as privacy concerns, location inaccuracies, and app errors, were not significant.

Negative associations with uninstallation include positive attitudes toward the app (OR: 0.091, 95\% CI 0.036-0.23) and positive beliefs in the app's utility (OR: 0.27, 95\% CI 0.11-0.63). Some demographic factors were found to be significantly associated with uninstalling the app. Nonacademic users are more likely to uninstall the app (OR: -.27, 95\% CI 0.11-0.63). Other factors, such as gender, age, technological level, and religion, were not significant.

\section{Discussion}
The COVID-19 pandemic has provided a rare opportunity to study how citizens' preferences matter for the effectiveness of public policies and are affected by the deployment of technological systems. Previous studies have looked at contact tracing apps in isolation, modeling the decisions of people under the assumption that there are no other systems that carry out contact tracing \cite{chan2021privacy,hassandoust2021individuals}. In contrast, we investigate a setting in which voluntary and involuntary systems are deployed simultaneously and demonstrate a possible spillover effect that arises from the existence of the involuntary system.

The relationship between utility perceptions of mass surveillance and the avoidance of contact tracing apps points to a mechanism that crowds out voluntary participation. What are the potential reasons behind crowding out in the context of COVID-19 contact tracing? We distinguish between two types of systems: "hands-on systems" that require consent and active participation versus "hands-off systems" that rely on ubiquitous mass surveillance. Hands-on systems are voluntary and are often operationalized by technology that the user needs to actively install and enable, such as contact tracing apps. Hands-off systems are involuntary and do not require citizens to carry out any action. We observe that while perceptions of app utility and mass surveillance utility are highly correlated, a multiple regression model shows that their impact on installation is drastically different.

\subsection{Crowding Out Mechanisms}

People's motivations may differ under hands-on and hands-off systems, not only because hands-off methods make participation redundant but also because their existence may affect people's preferences \cite{bowles2008policies}. It is essential to recognize that conceptualizing this problem requires extending the framework of motivational crowding out theory beyond its original premise. Unlike classic crowding-out scenarios, Israeli citizens were not offered a reward to use the mass surveillance system, nor was there a punishment involved in not using it. However, intrinsic motivations such as altruism are associated with installing COVID-19 contact tracing apps \cite{kaptchuk2020good,li2021makes,schmelz2021enforcement}. Therefore, we look at mechanisms of crowding out of social preferences that are discussed extensively in the literature \cite{bowles2008policies} and extend them with possible spillover monitoring mechanisms \cite{belot2016spillover,frey2001motivational} that are relevant to the unique aspects of surveillance.

First, implementing a mass surveillance system reveals the government's beliefs about the ability of citizens and the trust it has in them. Choosing a voluntary app may signal the government's confidence in citizens' responsibility \cite{sliwka2007trust}. In contrast, mass surveillance may signal the government's belief that people cannot be trusted to install the application and voluntarily share information in the case of detection of proximity events. Monitoring can lead to a lower level of trustworthiness by agents \cite{huang2006trust} and to lower productivity when workers retaliate for being distrusted \cite{belot2016spillover}. While we obtained these results in the context of workplace monitoring, it may be the case that similar reciprocity mechanisms can explain reactions to governmental mass surveillance.

Second, monitoring and enforcement may compromise personal autonomy \cite{ziegelmeyer2012hidden}. Unlike the contact tracing app, mass surveillance technology does not leave it to citizens to actively decide whether to notify health authorities about a contact event. Monitoring often backfires, with workers performing worse when they are monitored \cite{frey2001motivational,belot2016spillover}. Draper and Turow describe how mass surveillance leads to "digital resignation", people's feelings of futility in understanding and committing to action regarding their digital rights \cite{draper2019corporate}. Therefore, citizens might give up on autonomous technologies given the far-reaching restrictions of personal freedom, privacy, and autonomy.

Third, hands-on systems require some setup and incur usage costs. People need to download the app and install it, which may be difficult. Hands-off methods rid citizens of these costs, which may decrease the incentive to participate if they believe in the system's utility. Long-term usage incurs ongoing costs, such as battery drainage, which might hinder people's willingness to run the app, especially if an equivalent solution is available. Our findings point to the strong effect of concerns about the app's battery consumption in uninstalling the app.

Fourth, hands-on systems such as the contact tracing app require users to make moral decisions when installing, allowing access to Bluetooth or location services, which will require the user to make a judgment call on privacy. Additionally, it is up to the user to notify the health authorities when a proximity event is detected. The individual needs to make a difficult decision: inform the health authorities and act, which will probably result in quarantine? Or ignore the event? \cite{jamieson2021deciding} Given the uncertainty of the detection of the event and the fact that the contact tracing app does not provide information about the context of the event apart from the time of day, offloading the moral judgment to the mass surveillance system is understandable.

\subsection{Consequences of Crowding Out}
What are the implications of crowding out behaviors? Beyond the legal and ethical criticisms, mass surveillance systems go against the most fundamental principles of consent and control in fair information practices \cite{marciano2021israel}. It is clear why decision-makers might prefer a hands-off approach to contact tracing: it does not require authorities to market and convince people to download the app. It might be easier to collect information from the system, and it might be faster to deploy the system. However, understanding crowding out mechanisms may also lead to a closer evaluation of the effectiveness of mass surveillance systems in light of spillover effects. Crowding out effects may be harmful to the overall performance of contact tracing if there is an imbalance in the effectiveness of the two systems. When the hands-off system is less effective than the hands-on system, but it crowds out users from the hands-on approach, then it may hamper the overall objective of the two systems.

In the case of Israel, we have evidence that the mass surveillance system was not efficient in detecting proximity events, and therefore deploying it may have hurt the Israeli contact tracing efforts. The mass surveillance project in Israel revealed the limitations of the system. According to the State Comptroller's October 27 report, 3.5\% to 4.7\% of those told to quarantine based on the mass surveillance methods contracted the coronavirus, compared with 24\% of those told to quarantine by an epidemiological investigation team \cite{comptroller2020operating}. The false positively rate of the mass surveillance tool was also high: the system unnecessarily sent into quarantine three to eight times as many people than the manual epidemiological processes. 60\% of the appeals against self-quarantine orders due to contact with a verified coronavirus patient were granted \cite{lis2020about}.
Our findings show that the people's trust in the mass surveillance system is very limited. Our survey shows that the acceptance of behaviors such as avoiding carrying phones is high. If many people refrain from carrying their phones, the overall accuracy of the system may deteriorate. The low levels of trust regarding the deletion of data after the pandemic are another reason that may push people to limit their cell phone use.

How much can we generalize from the Israeli case? Several countries have deployed involuntary contact tracing systems in addition to Israel, including China, South Korea, and Taiwan \cite{Altshuler2020Digital}. Several other countries have considered various forms of centralized contact tracing technologies \cite{abbas2020covid}. At the same time, many other countries have multiple contact tracing technologies working in parallel, with private companies and other local governments each requiring different apps \cite{machida2021survey}. Our study is relevant to identify and analyze the possible spillover effects in the deployment of multiple technologies. More broadly, our findings are also relevant to situations in which hands-off and hands-on methods compete for similar objectives, situations that are becoming more widespread as artificial intelligence systems are becoming more widespread in tracking and interpreting people's behavior.

\subsection{The Larger Scope of Contact Tracing App Adoption}

Our model of users' attitudes toward contact tracing app installation and uninstallation has implications beyond our main research question about crowding-out effects. Our findings show that privacy perceptions are essential to the installation of contact tracing apps. Our results support previous studies in other parts of the world about the connection between deeper privacy concerns and lower acceptance of contact tracing apps \cite{chan2021privacy,altmann2020acceptability}. Our findings show that people's attitudes toward contact tracing technologies are shaped by their perceptions of the health benefits that they offer to them individually or to the people around them, supporting existing studies (e.g., \cite{altmann2020acceptability,chan2021privacy,fox2021exploring}).

We found some demographic differences in attitudes toward contact tracing apps. Males are generally more inclined to install the apps, as was found in several other studies \cite{fox2021exploring, kim2021covid, kaptchuk2020good}. This gender difference aligns with past research that showed that women are more privacy sensitive than men \cite{redmiles2018net}. We also found higher levels of installation for people with graduate degrees, a phenomenon also reported in several other studies \cite{machida2021survey,zhang2020americans}. Unlike other studies \cite{altmann2020acceptability,machida2021survey}, we did not see strong effects of age on installation rates. The reason might be that our survey asked about actual installations rather than the willingness to install, with some additional technical barriers for older users.

Our results also demonstrate that people find it difficult to assess the privacy risks associated with different types of contact tracing technology. The privacy concerns associated with mass surveillance and with the privacy-preserving contact tracing apps were too small to be significant. These results support a previous vignette study by Li et al. that shows that people cannot easily distinguish between centralized and privacy-preserving contact tracing apps \cite{li2021makes}.

Our study analyzed uninstallations of the COVID-19 contact tracing app. In France, a recent study found that among their participants, most uninstallations were due to negative perceptions of the app's usefulness, forgetting to activate the Bluetooth, and battery draining \cite{montagni2021health}. In our study, we have found that people uninstall the app if their attitudes toward it are negative, if the attitudes toward mass surveillance are positive, if they believe that it will drain their battery, and if they were not found to have COVID. Men uninstall it more than women. We see that the model for installation has some resemblance to the model for uninstallation but with some considerable differences. Privacy concerns do not play a role in making decisions about uninstallations, perhaps because people already made the decision to share information with the app.

\subsection{Implications for Policy and Design}

Our analysis exemplifies the complexity of mass surveillance in general and in contact tracing technologies in particular. It adds to the growing literature that points to behavioral spillover effects in COVID-19 when involuntary measures are deployed, such as social distancing mandates \cite{yan2021measuring} or vaccine mandates \cite{schmelz2021overcoming}.

First, deploying an obligatory mass surveillance system may have a substantial negative impact on the voluntary installation of contact tracing apps and may be harmful if involuntary methods are not practical. Given the limited ability of citizens to understand digital technologies, enforcing involuntary systems can confuse citizens and discourage them from adopting more complicated technologies. Deploying mass surveillance technologies might imply costs, including reduced trust in health authorities, growing norms of avoiding surveillance (such as leaving the phone at home), and reduced confidence in privacy-preserving solutions. We see that Israel's decision to rely on involuntary mass surveillance did not contain the coronavirus pandemic. Privacy concerns and an erosion of trust have led people to engage in subversive behaviors, such as leaving their phones at home \cite{JPphone2020}.

Second, an essential lesson in COVID-19 contact tracing is to reduce privacy concerns. We see that installation is strongly associated with lower privacy concerns and minimizing these concerns by developing privacy-preserving technologies can result in higher installation rates. Active measures to enhance the privacy of contact tracing technology could reduce the friction of installing these apps and increase adoption. However, our findings point to the challenge of communicating privacy-preserving technologies; participants had similar privacy concerns about both the mass surveillance and the contact tracing apps, while in practice, the cellular tracking technology violates the public's privacy more extremely. Therefore, we suggest communicating the app's privacy benefits in a prominent way to motivate the public to download the app.

To ethically design these applications, designers need to balance the benefits and harms of data collection and ensure that the benefits of contact tracing are distributed fairly \cite{parker2020ethics}. Privacy-by-design principles were applied to the design and deployment of CTT, including data minimization, consent, proper oversight, and due processes. Given the urgent and unexpected nature of the pandemic, the technology can come with an expiration date and proper supervision that would ensure that the extraordinary measures of contact tracing will not be used for any purposes other than restricting the spread of the pandemic.

\section{Conclusions}
We modeled users' attitudes toward contact tracing app installation and uninstallation and showed that they are related to attitudes toward mass surveillance contact tracing. In this study, based on a representative survey of Israelis, we used logistic regression models to investigate how people's perceptions about the mass surveillance COVID-19 apparatus and the contact tracing app relate to installing and uninstalling the contact tracing app. Our findings suggest that the effects of mass surveillance are meaningful even when controlling for attitudes toward contact tracing privacy, utility, and demographics.

\bibliographystyle{abbrvnat}
\bibliography{covid19}

\appendix

\section{Detailed Description of Contact Tracing Technologies in Israel}

\begin{figure}[ht]
  \centering
  \includegraphics[width=0.7\textwidth]{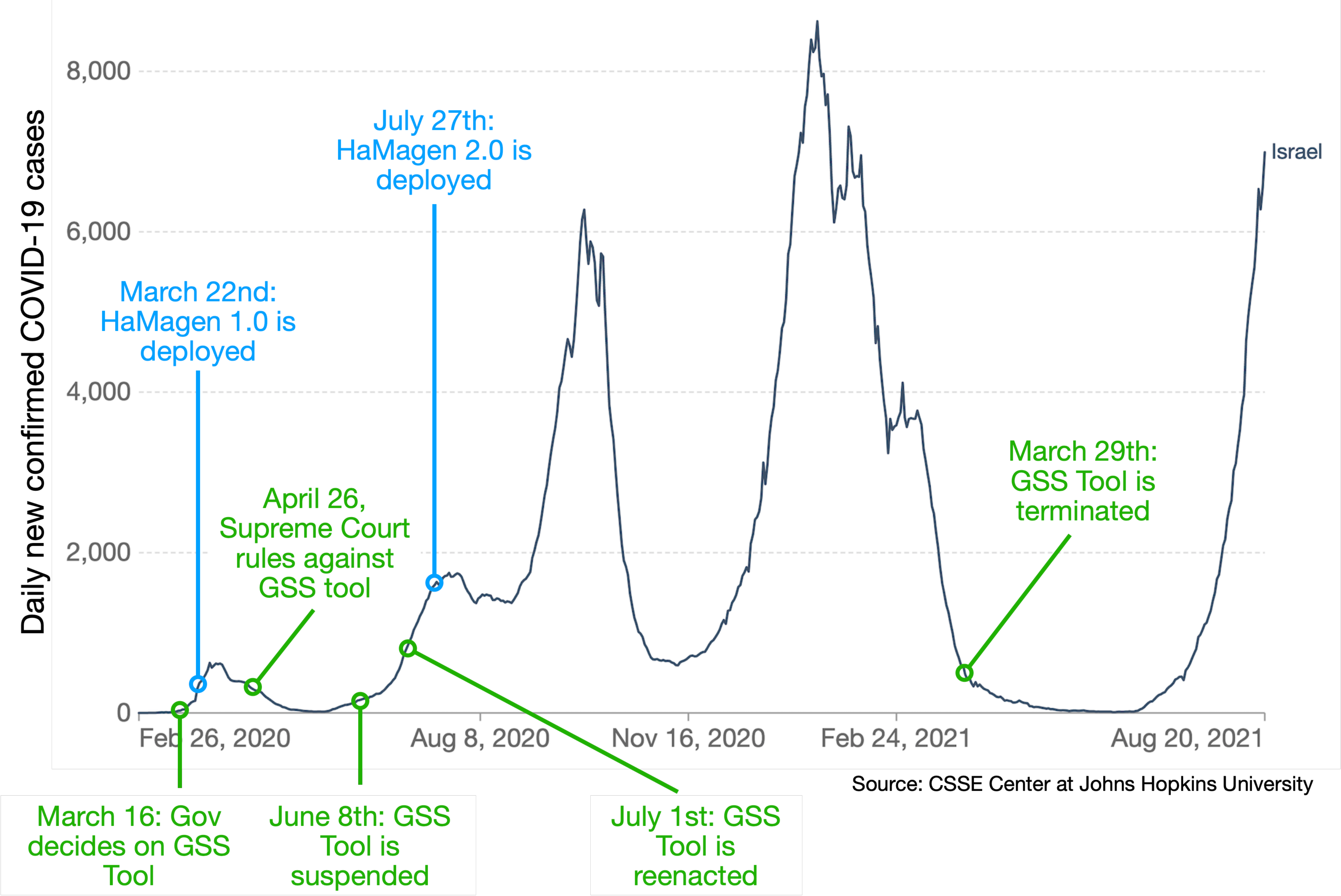}
  \caption{A timeline of the implementation of Contact Tracing Technologies in Israel}
  \label{fig:timeline}
\end{figure}

In Israel, two contact tracing technologies were operable in the duration of the study: \emph{HaMagen} ("the Shield", in Hebrew), a contact tracing application that was developed by the Ministry of Health, and centralized cellular tracking technology that is operated by Israel's General Security Services (GSS), dubbed "The Tool." The timeline of the deployment of the two technologies is depicted in Figure \ref{fig:timeline}. 

HaMagen ("the Shield", in Hebrew), is a contact tracing application developed by the Ministry of Health. The first version, HaMagen 1.0, was based on the ongoing local storage of users' location data and local matching with official data about infected people's whereabouts. The government provided links to downloading the app were on the Health Ministry Website, but was not widely promoted in the media. Even with the limited exposure, about 1.5 million people have downloaded it and 400,000 people have uninstalled it \cite{globes2020tracking}. 

\begin{figure}[ht]
\centering
\begin{minipage}[b]{0.45\linewidth}
\includegraphics[width=\textwidth]{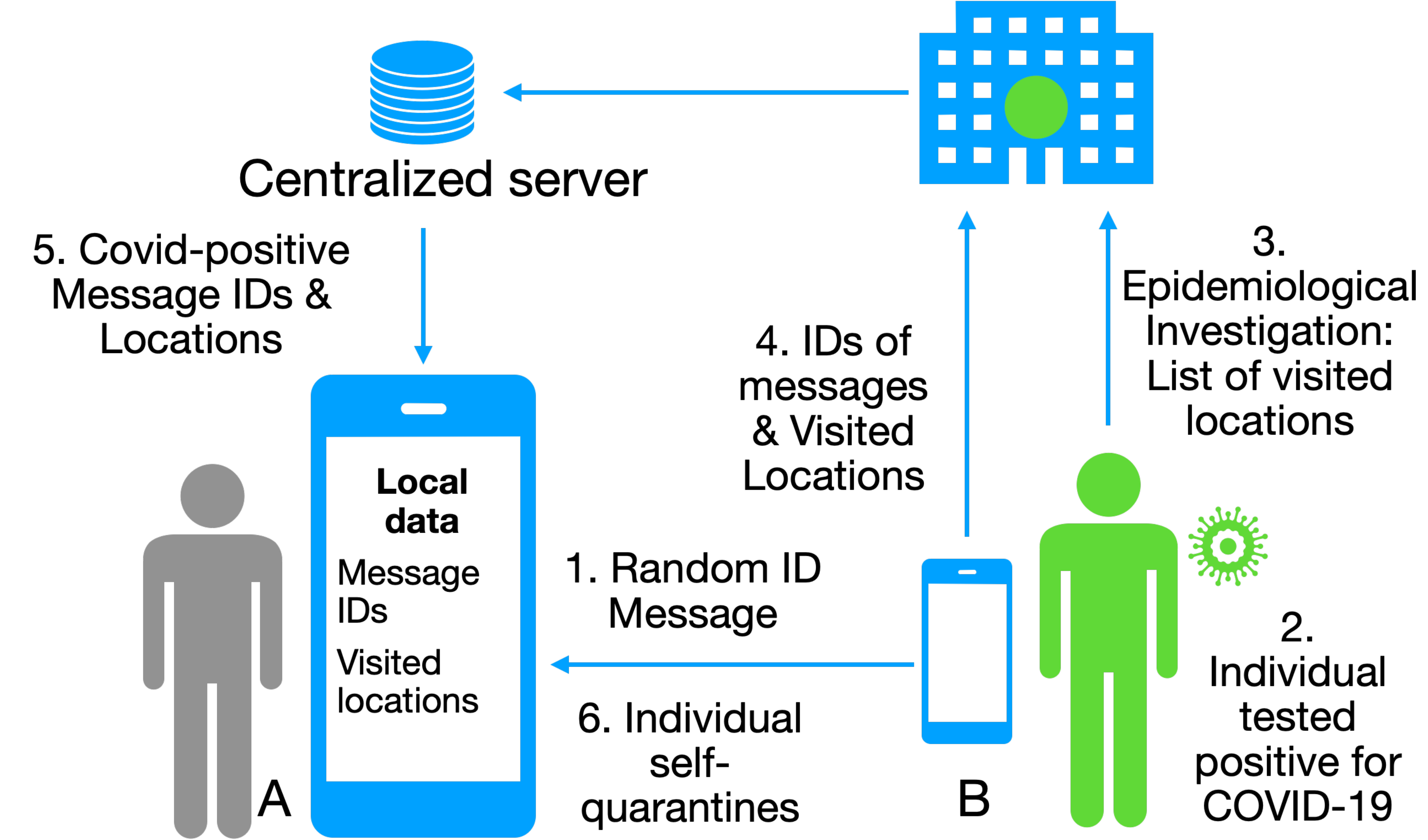}
\caption{Hamagen Architecture}
\label{fig:hamagen_arch}
\end{minipage}
\quad
\begin{minipage}[b]{0.45\linewidth}
\includegraphics[width=\textwidth]{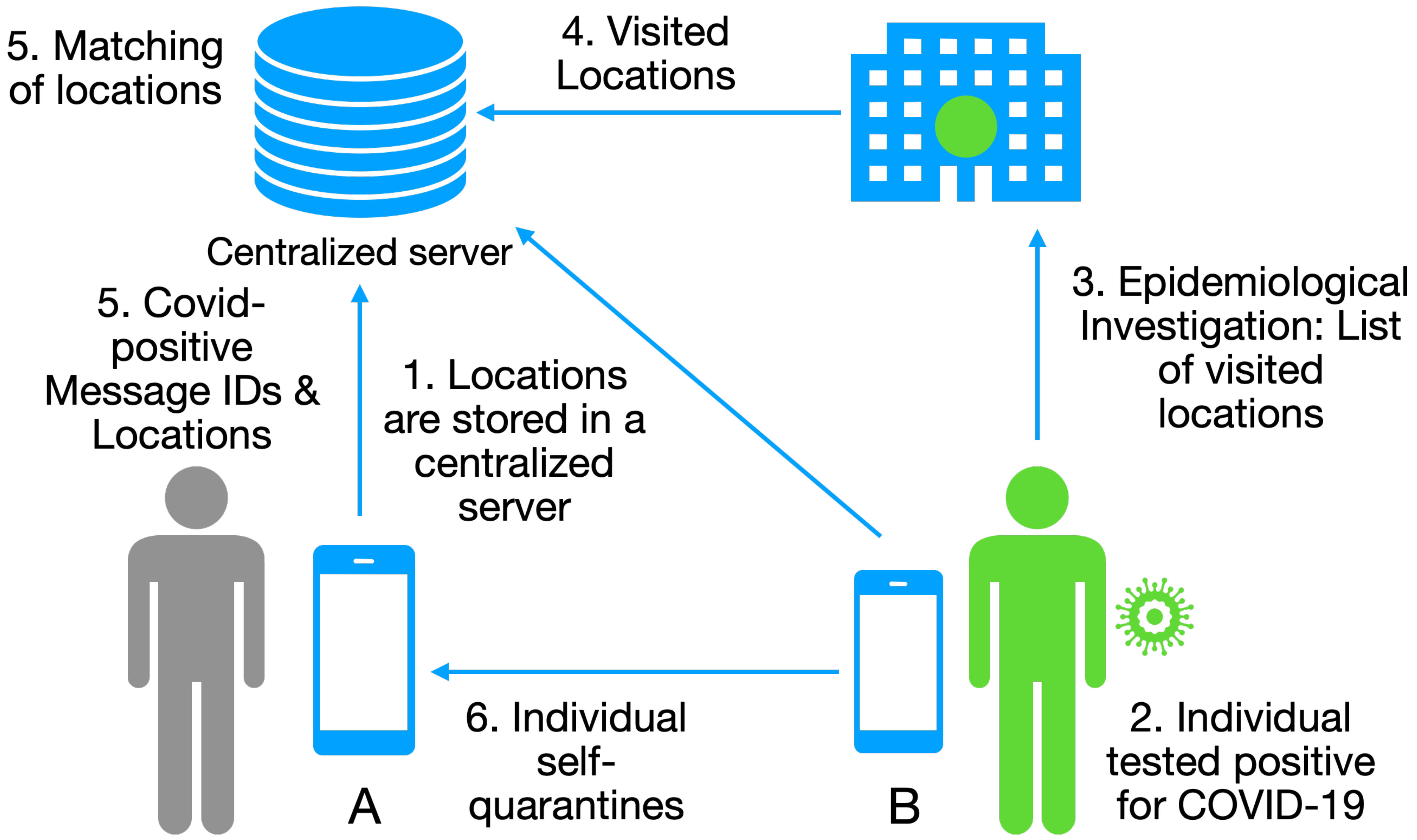}
\caption{The Tool Architecture}
\label{fig:tool_arch}
\end{minipage}
\end{figure}

HaMagen mobile application collected information about the visited locations using the mobile phone's GPS and Wi-Fi positioning capabilities \cite{altshuler2020contact}. Beginning with the second version of HaMagen, the app also received messages from nearby phones through BLE. These messages contained randomly assigned IDs and theoretically cannot be used to identify the nearby telephone. When an individual is identified as COVID-19 positive, they are briefed by an epidemiological investigation team (see Figure \ref{fig:hamagen_arch}). 

The locations users visited within the past two weeks were fed into a simple centralized server. If the individual had the HaMagen app installed, health authorities could upload the locations and BLE messages to the server. Each app regularly retrieves the list of places and message IDs. If there was a match with the places or the messages received from a COVID-19 positive person, the user was notified and was asked to contact the health authorities.

The GSS Tool was based on centralized cellular tracking operated by Israel's General Security Services (GSS). The technology was based on a framework that tracks all the cellular phones running in Israel through the cellular companies' data centers. According to news sources, it routinely collects information from cellular companies and identifies the location of all phones through cellular antenna triangulation and GPS data. Regularly, the Tool is used according to authorization from a court order, but on March 16, the Israeli government authorized the use of this technology for contact tracing \cite{nytimes2020track}. 

\begin{figure}[ht]
  \centering
  \includegraphics[width=0.6\textwidth]{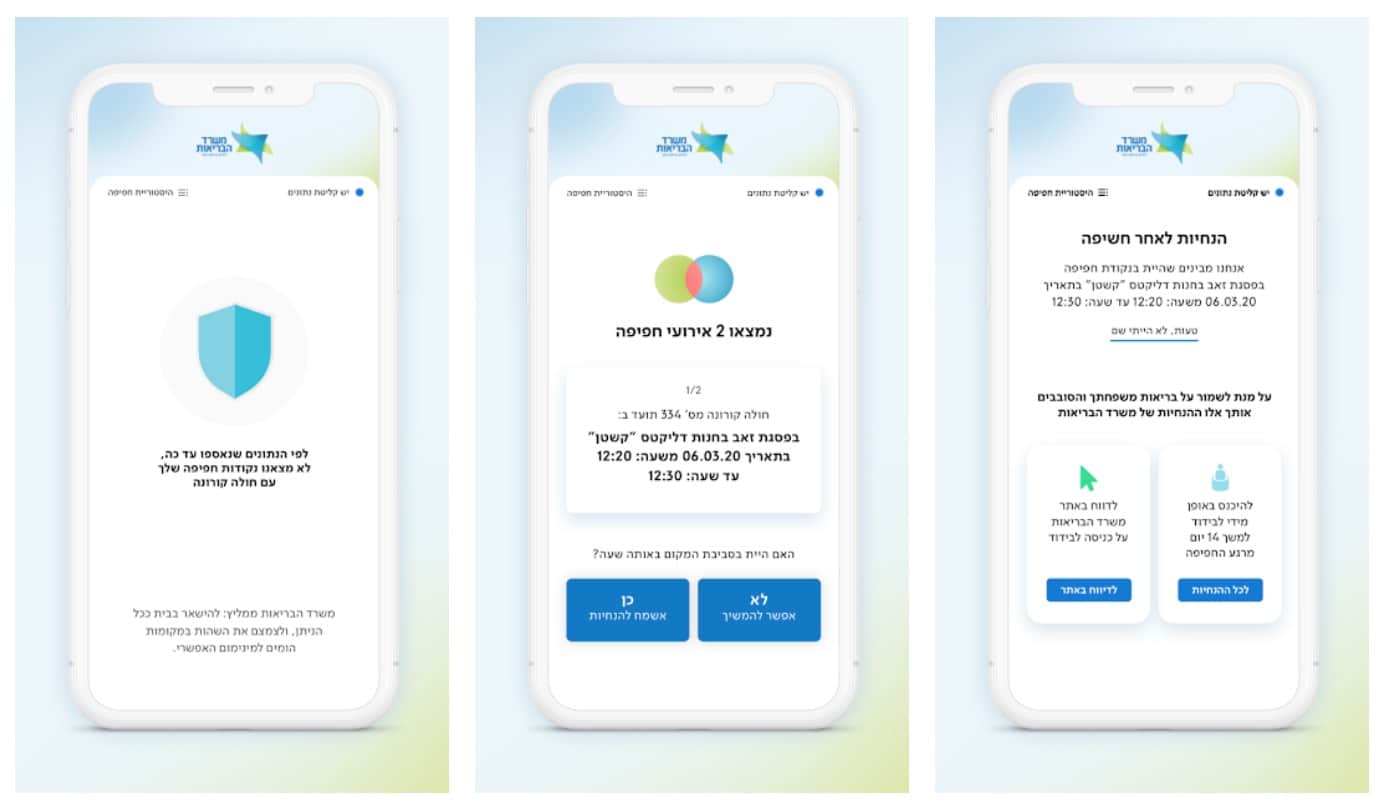}
  \caption{Screenshots from the HaMagen application)}
  \label{fig:hamagen_app}
\end{figure}

The Tool traces contacts with constant location tracking carried out through Israel's cellular companies (see Figure \ref{fig:tool_arch}). Cell phone's location is tracked using a mixture of GPS locations transmitted through cellular protocols and cellular antenna triangulation \cite{nytimes2020track}. When individuals identified as COVID-19 positive, they were briefed by the epidemiological investigation teams, and the locations they visited during the previous two weeks were fed into The Tool. The system analyzed the location data and pinpointed individuals who were close to the COVID-positive case. Contact details for individuals identified by The Tool are then sent to the health authorities, which notify them via text message that they have to self-quarantine. The system did not let people know the location or the exact time of their interaction with the infected individual.

\begin{figure}[ht]
  \centering
  \includegraphics[width=0.4\textwidth]{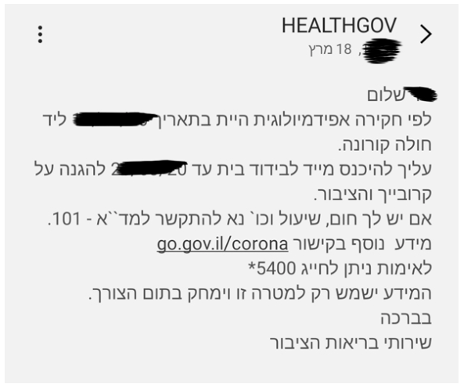}
  \caption{An example for a text message received from the Ministry of Health after being tracked by the GSS Tool. Translated text from Hebrew: The recipient is informed that, according to an epidemiological investigation, they have been close to a verified Coronavirus patient and must enter home quarantine. )}
  \label{fig:sms}
\end{figure}

\section{Method Details}
The survey instrument included several questions about voluntary and involuntary contact tracing technologies. The complete questionnaire and all the items are available in Table \ref{tab:items}. The participants were first presented with general questions about their attitudes toward the COVID-19 crisis (GA1-7) and some risk factors (RF1-6). Afterward, they were presented with questions about the contact tracing app and mass surveillance technology. For each technology, we asked the participants to read a short excerpt from a daily newspaper that describes the technology and answer several questions about their attitudes toward it. We have randomized the order of the two technologies.

To assess the basic attitudes about the two technologies, we used four identical questions: app utility questions were based on the World Health Organization (WHO) guidelines for COVID-19 survey tools \cite{who2020survey} and asked about the perceived utility of the app. Two other questions measure the perceived privacy concerns of participants based on a validated privacy index that measures concern and sensitivity \cite{kumaraguru2005privacy}.

Several questions were specific to the contact tracing app. First, we asked whether the participants had installed or uninstalled the app. We also asked several specific questions regarding general attitudes toward the contact tracing app (AT1 and AT2). We asked the participants who had uninstalled the application to respond to three specific potential reasons: 1) "The app does not always recognize my location accurately"; 2) "I have encountered errors in the app"; and 3) "The app wastes my battery".

We collected demographic information about gender, age, religion, level of observance, education, and income. The questions and summary statistics are available in Table \ref{tab:demographics}. In addition, to assess \textit{technical abilities}, we have used a verified scale by Hargittai (2005), which measures people's familiarity with digital technologies \cite{hargittai2005survey}.

\begin{table}[ht]
\begin{tabular}{p{2cm} p{1cm} p{1cm} p{8cm} p{1cm} p{1cm}}
\hline
Construct  & $\alpha$ & Item & Text                                                                                  & Mean & STD  \\ \hline
App Attitude        & 0.72        & AT1  & I will recommend installing the app to my friends and family                               & 2.75 & 1.22 \\
                            & & AT2  & People who enter malls or public transport should be required   to install the HaMagen app & 2.82 & 1.43 \\
App Utility         & 0.84        & AU1  & The app will help reduce the    spread of the coronavirus                                  & 3.12 & 1.13 \\
                            & & AU2  & The app will reduce the chances that I will catch the   coronavirus                        & 2.82 & 1.16 \\
App Privacy                 &  0.78 &AP1  & I am worried of the information the app can collect on me                                  & 3.17 & 1.35 \\
                            &  & AP2  & The app can collect sensitive information                                                  & 3.62 & 1.21 \\ \hline
Surveillance Utility & 0.82 & SU1 & The technology will help reduce the  spread of the novel coronavirus                       & 3.21 & 1.09 \\
                        &    & SU2 & The technology will reduce the chances that I will get   infected by the coronavirus       & 2.89 & 1.10 \\
Surveillance Privacy   & 0.79 & SP2 & I am worried of the information the technology can collect on me                        & 3.20 & 1.35 \\
                          &  & SP3 & The technology can collect sensitive information                                           & 3.70 & 1.14 \\
                            &  \\
Trust Delete   & & STD & I trust that all the data will be deleted after the end of the   coronavirus crisis        & 2.46 & 1.22 \\

Understand Leave   & & SUL & I can understand people who leave their phone at home to avoid   cellular tracking         & 2.92 & 1.32 \\
                        \hline
COVID-19 Concerns   & 0.77 & CC1  & The novel coronavirus threatens the population of Israel                                    & 3.13 & 1.17 \\
                          &  & CC2  & The novel coronavirus threatens my health                                                  & 2.86 & 1.15 \\
                         &   & CC3  & I am worried that people I know will get infected by the novel   coronavirus               & 3.61 & 1.16 \\
                         &   & CC4  & I am worried to get infected by the novel coronavirus                                      & 3.03 & 1.24 \\
Trust &   & GA5  & I have trust in the professional authorities that lead the   handling of the coronavirus   & 3.25 & 1.09 \\
                            
Compliance & & GA6  & I strictly follow the instructions of the health ministry to   fight the coronavirus          & 4.18 & 0.83 \\

Financial hurt &  & GA7  & The coronavirus crisis is hurting me financially                                           & 2.92 & 1.29 \\ \hline
\end{tabular}
\caption{A list of all constructs and items questionnaire. Cronbach’s alpha is the scale reliability coefficient for  internal consistency. The text was translated from Hebrew.}
\label{tab:items}
\end{table}

\begin{table}[ht]
\begin{tabular}{p{2cm} p{2cm} p{8cm} p{1cm} p{1cm}}
\hline
Construct     & Item & Item Text                                                            & No  & Yes \\ \hline
Risk factors & RF1  & Do you work in a healthcare clinic or a hospital                     & 487 & 32  \\
             & RF2  & Are you in a special risk of the coronavirus                         & 409 & 110 \\
             & RF3  & Do you have relatives who are in special risk of the virus           & 84  & 435 \\
             & RF4  & Do you know a person that was tested positive for the   coronavirus? & 373 & 146 \\
             & RF5  & Were you ever tested positive for the coronavirus?                   & 509 & 10  \\
             & RF6  & Are you or were you in quarantine?                                   & 475 & 44  \\ \hline
\end{tabular}
\caption{A list of all binary items. The text was translated from Hebrew.}
\label{tab:binary_items}

\end{table}

A pilot survey of 50 participants was conducted beforehand to assess the quality of the questionnaire and initial effect sizes. As a result of the pilot study, we made several minor changes to the questionnaire. As almost all participants had heard about the contact tracing technologies, we removed questions about familiarity with the technologies. Furthermore, we have removed some questions that were too general and unintelligible. The survey was administered through an online panel by a commercial firm that carries out Internet panel surveys. We used stratified quota sampling to approximate the marginal distributions of vital demographic characteristics in Israel: religion, gender, and age.

Our data analysis is based on two logistic regression models: to predict the likelihood of participants installing the contact tracing app and the likelihood of participants who already installed the app to uninstall it. To identify critical determinants of contact tracing app installations, we first computed a categorical variable indicating whether the app was installed:
\begin{equation}
Install = 
\begin{cases}
    0, & \text{Not installed: if the app was not installed}  \\
    1, & \text{Installed: if the app is currently installed} \\
    2, & \text{Uninstalled: if the app was removed from the phone}
    \end{cases}
\end{equation}

We fitted the logistic regression models on subsets of the whole dataset. We fitted the "install" model to participants who had not installed the app, while we fitted the "uninstall" model to participants who either had installed or uninstalled the app. We opted to use two models instead of a categorical model (such as a multinomial logistic regression) because three questions were presented only for people who had installed the app.

Data verification focused on ensuring that the data met the thresholds required for our analysis \cite{hair2010multivariate}. To examine multicollinearity, the variance inflation factor (VIF) was calculated for all variables. Two factors (app utility and mass surveillance utility) were above the threshold \cite{hair2010multivariate}, so they were standardized using the centering method \cite{iacobucci2016mean}. After centering, the VIFs of all variables were below the threshold. The proposed constructs had Cronbach’s alpha values above 0.75, which point to good reliability (all Cronbach alpha values are presented in Appendix B). All data analysis tasks were carried out in STATA (ver. 17). 

\begin{table*}[ht]
\centering
\begin{tabular}{p{0.8cm}p{0.20cm}p{0.02cm} p{0.9cm}p{0.20cm}p{0.02cm} p{0.8cm}p{0.20cm}p{0.02cm} p{1.2cm}p{0.20cm}p{0.02cm} p{1.2cm}p{0.2cm}p{0.02cm} p{1.0cm}p{0.2cm}l}
\hline
\textbf{Gender} & \textbf{} & \textbf{} & \textbf{Age} & \textbf{} & \textbf{} & \textbf{Religion} & \textbf{} & \textbf{} & \multicolumn{2}{l}{\textbf{Observance}} & \textbf{} & \textbf{Education} & \textbf{} & \textbf{} & \multicolumn{2}{l}{\textbf{Income}} & \textbf{} \\ \hline
Male            & 239       &           & 18-19        & 10        &           & Jewish            & 426       &           & Secular                    & 190               &           & Elementary         & 3         &           & \multicolumn{2}{l}{Strong Below Average}  & 141       \\
Female          & 276       &           & 20-29        & 158       &           & Muslim            & 55        &           & Traditional                & 139               &           & Highschool         & 112       &           & \multicolumn{2}{l}{Below Average}           & 165       \\
Other           & 4         &           & 30-39        & 139       &           & Druze             & 22        &           & Religious                  & 93                &           & Non                & 138       &           & \multicolumn{2}{l}{Average}                 & 138       \\
                &           &           & 40-49        & 81        &           & Christian         & 16        &           & Orthodox                   & 57                &           & Bachelor           & 184       &           & \multicolumn{2}{l}{Above Average}           & 61        \\
                &           &           & 50-59        & 76        &           &                   &           &           &                            &                   &           & Graduate           & 81        &           & \multicolumn{2}{l}{Strong Above Average}  & 14        \\
                &           &           & 60-69        & 52        &           &                   &           &           &                            &                   &           &                    &           &           &                      &                      &           \\
                &           &           & 70-80        & 3         &           &                   &           &           &                            &                   &           &                    &           &           &                      &                      &           \\ \hline
\end{tabular}
\caption{Sample demographic characteristics}
\label{tab:demographics}
\end{table*}

\section{Summary Statistics of Participants' Answers}

\begin{figure}[p]
  \centering
  \includegraphics[width=0.7\textwidth]{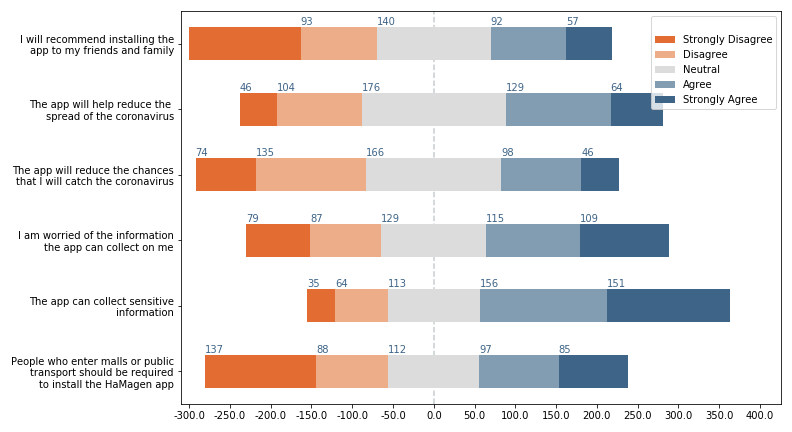}
  \caption{Overall attitudes regarding the HaMagen app (on a 5-point agreement Likert scale)}
  \label{fig:hamagen_attitudes}
\end{figure}

We have analyzed the basic attitudes toward the HaMagen contact tracing app. Figure \ref{fig:hamagen_attitudes} shows the results of our participants' self-reported answers. Overall, people express negative attitudes regarding the app. Only 28\% will tend to recommend their friends and family install it (versus 43\% that would not). A somewhat higher proportion have a positive attitude towards making it mandatory for people entering malls or public transportation (35\% versus 43\%). Only 27\% think it will reduce their chances of contracting the coronavirus, and only 32\% believe it will reduce the spread of the virus. Privacy concerns are prevalent. 59\% feel that it collects sensitive information (versus 19\% that disagree), and 43\% are worried about privacy (versus 32\% that disagree).

\begin{figure}[p]
  \centering
  \includegraphics[width=0.7\textwidth]{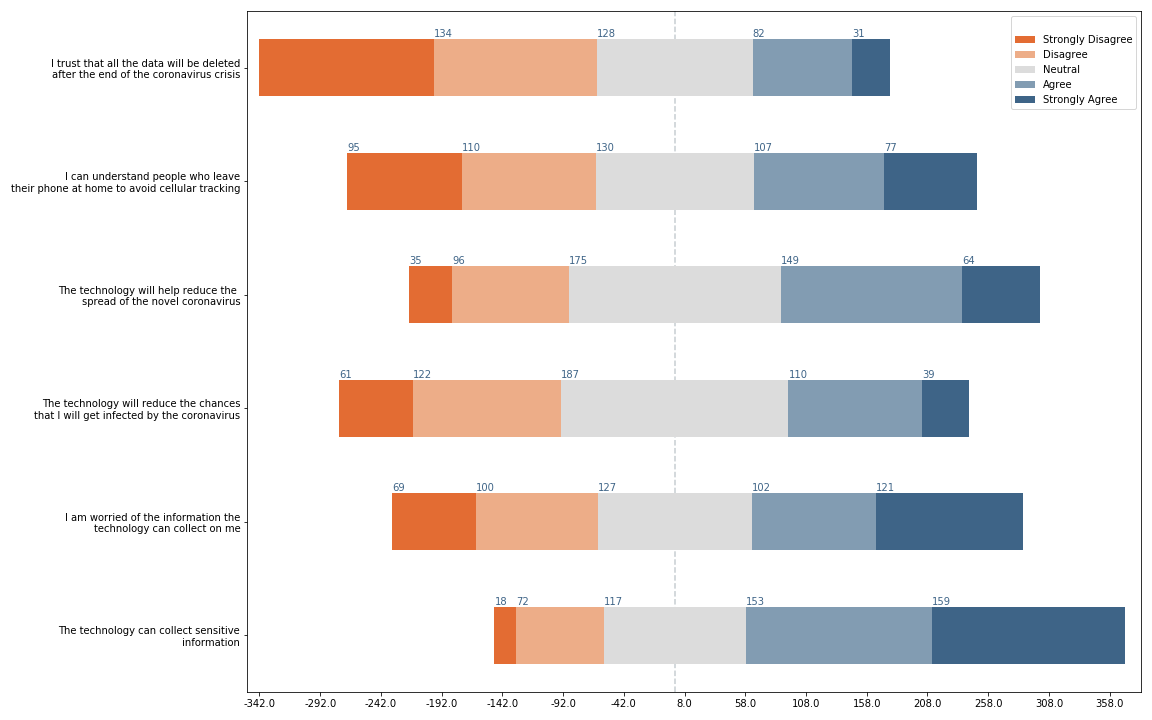}
  \caption{Overall attitudes regarding the GSS Tool cellular tracking (on a 5-point agreement Likert scale)}
  \label{fig:cellular_attitudes}
\end{figure}

How do Israelis view the involuntary cellular coronavirus contact tracing technology operated by Israel's GSS? The summary of the answers is presented in Figure \ref{fig:cellular_attitudes}. Our survey shows that most Israelis do not trust the government to delete the data after the crisis is over (53\% disagree  versus 21\% who agree). About 35\% of our participants are sympathetic to people leaving their phones at home to avoid being tracked (versus 39\% are have an unfavorable view). About 60\% agree that cellular tracking can collect sensitive information (versus 17\% who disagree). About 42\% report privacy concerns because of the cellular tracking technology, versus 32\% who disagree. 

\begin{figure}[p]
  \centering
  \includegraphics[width=0.7\textwidth]{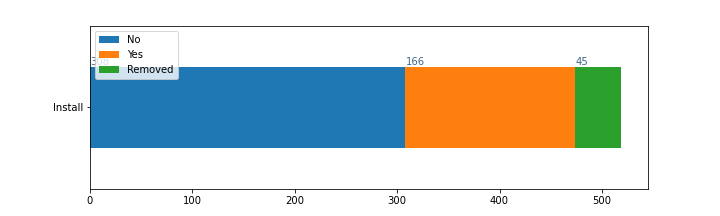}
  \caption{Proportions of participants who installed and uninstalled the contact tracing app}
  \label{fig:hamagen_install}
\end{figure}

We compared the attitudes towards the contact tracing application to the centralized cellular tracing contact tracing technology. Overall, we did not find statistically significant differences in the approaches towards privacy between the two architectures. The medians and variances visually look very similar. A Wilcoxon sum test did not find significant differences (W=17499.0, $p=0.15$). The differences between the perceived utility are statistically significant, but the effect size is rather small. The median utility is identical, but more participants believe cellular tracing offers more utility (Wilcoxon sum test, W=18579.5, p=0.018). 

\pagebreak 

\section{Detailed Logistic Regressions Results}

\begin{table}[h]
\small
\begin{tabular}{llllllll}
\hline
Install                  & Odds. & St.Err. & t-value & p-value & {[}95\% Conf & Interval{]} & Sig \\ \hline
App Attitude            & 3.923 & .592    & 9.05    & 0       & 2.918        & 5.274       & *** \\
Surveillance Attitude   & .547  & .107    & -3.08   & .002    & .372         & .803        & *** \\
App Privacy             & .758  & .087    & -2.40   & .016    & .605         & .95         & **  \\
App Utility             & 1.05  & .176    & 0.29    & .772    & .756         & 1.458       &     \\
Tech Level              & 1.178 & .139    & 1.39    & .165    & .935         & 1.485       &     \\
Gender: Male            & 2.041 & .54     & 2.69    & .007    & 1.215        & 3.429       & *** \\
Age: 18-20              & 5.394 & 4.407   & 2.06    & .039    & 1.088        & 26.752      & **  \\
Age: 30s                & 1.159 & .384    & 0.44    & .656    & .605         & 2.218       &     \\
Age: 40s                & 1.313 & .494    & 0.72    & .469    & .628         & 2.745       &     \\
Age: 50s                & 1.177 & .479    & 0.40    & .688    & .53          & 2.613       &     \\
Age: 60s                & 1.149 & .517    & 0.31    & .757    & .476         & 2.773       &     \\
Education: Graduate     & 2.022 & .748    & 1.90    & .057    & .979         & 4.173       & *   \\
Education: Highschool   & .682  & .236    & -1.11   & .269    & .346         & 1.344       &     \\
Education: Non Academic & .677  & .218    & -1.21   & .227    & .36          & 1.274       &     \\
Christian               & 1.751 & 1.267   & 0.77    & .439    & .424         & 7.228       &     \\
Druze                   & 1.237 & .837    & 0.31    & .753    & .328         & 4.663       &     \\
Jewish                  & 1.93  & .809    & 1.57    & .117    & .849         & 4.388       &     \\
Constant                & .025  & .021    & -4.27   & 0       & .005         & .135        & *** \\ \hline
\end{tabular}
\caption{The association between installing the COVID-19 contact tracing app, attitudes, and demographic factors (n = 471). *** p<.01, ** p<.05, * p<.1.}
\label{tab:install}
\end{table}

\begin{table}[h]
\small
\begin{tabular}{llllllll}
\hline
Uninstall      & Odds. & St.Err. & t-value & p-value & [95\% Conf & Interval & Sig \\ \hline
App Attitude            & .091           & .043             & -5.06            & 0                & .036                  & .23                  & ***          \\
Surveillance Attitude   & 8.574          & 4.838            & 3.81             & 0                & 2.838                 & 25.909               & ***          \\
App Privacy             & 1.362          & .369             & 1.14             & .253             & .802                  & 2.315                &              \\
App Utility             & .274           & .116             & -3.05            & .002             & .119                  & .63                  & ***          \\
COVID Concern           & 1.19           & .347             & 0.60             & .55              & .673                  & 2.106                &              \\
Battery Concerns        & 2.232          & .632             & 2.83             & .005             & 1.281                 & 3.889                & ***          \\
Wrong Location          & .581           & .17              & -1.86            & .063             & .328                  & 1.03                 & *            \\
App Errors              & .838           & .205             & -0.72            & .47              & .518                  & 1.354                &              \\
Tech Level              & 1.018          & .271             & 0.07             & .947             & .604                  & 1.715                &              \\
Gender: Male            & .306           & .196             & -1.85            & .064             & .087                  & 1.071                & *            \\
Age: 20s                & 14.8           & 31.943           & 1.25             & .212             & .215                  & 1017.382             &              \\
Age: 30s                & 4.104          & 8.756            & 0.66             & .508             & .063                  & 268.719              &              \\
Age: 40s                & 4.943          & 10.734           & 0.74             & .462             & .07                   & 348.734              &              \\
Age: 50s                & 2.889          & 6.367            & 0.48             & .63              & .038                  & 217.023              &              \\
Age: 60s                & 13.273         & 29.733           & 1.15             & .248             & .164                  & 1071.105             &              \\
Education: Graduate     & .992           & .878             & -0.01            & .993             & .175                  & 5.621                &              \\
Education: Highschool   & 2.97           & 2.552            & 1.27             & .205             & .551                  & 15.998               &              \\
Education: Non Academic & 7.286          & 5.441            & 2.66             & .008             & 1.686                 & 31.485               & ***          \\
Christian               & 13.324         & 28.418           & 1.21             & .225             & .204                  & 871.125              &              \\
Druze                   & 3.408          & 5.547            & 0.75             & .451             & .14                   & 82.795               &              \\
Jewish                  & 4.75           & 6.481            & 1.14             & .253             & .328                  & 68.862               &              \\
COVID Positive          & .006           & .012             & -2.52            & .012             & 0                     & .322                 & **           \\
Income: Above Average   & 10.645         & 11.153           & 2.26             & .024             & 1.366                 & 82.972               & **           \\
Income: Average         & 4.56           & 3.812            & 1.81             & .07              & .886                  & 23.473               & *            \\
Income: Below Average   & .858           & .677             & -0.19            & .846             & .183                  & 4.025                &              \\
Income: High            & 8.479          & 15.344           & 1.18             & .238             & .244                  & 294.321              &              \\
Constant                & .063           & .18              & -0.96            & .335             & 0                     & 17.465               &              \\ \hline
\end{tabular}
\caption{The association between uninstalling the COVID-19 contact tracing app, attitudes, and demographic factors (n = 471). *** p<.01, ** p<.05, * p<.1.}
\label{tab:uninstall}
\end{table}

\begin{figure}[p]
  \centering
  \includegraphics[width=0.7\linewidth]{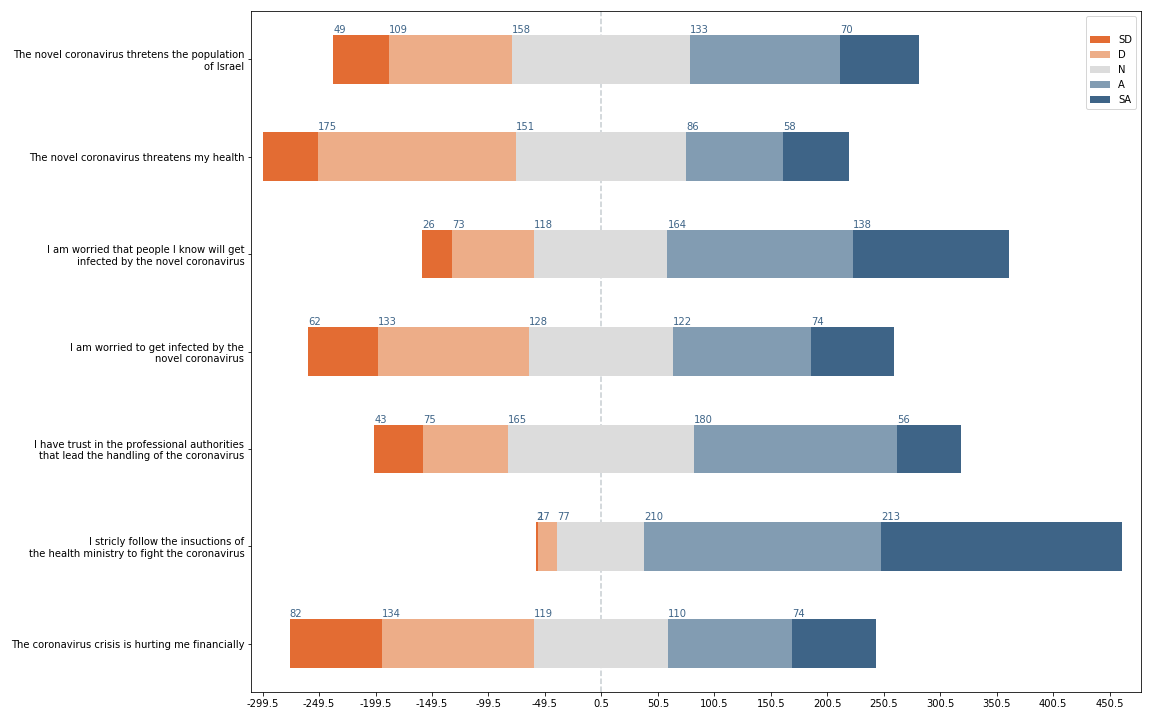}
  \caption{The distribution of the general attitudes (on a 5-point agreement Likert scale) }
  \label{fig:general}
\end{figure}

\begin{figure}[p]
  \centering
  \includegraphics[width=0.7\linewidth]{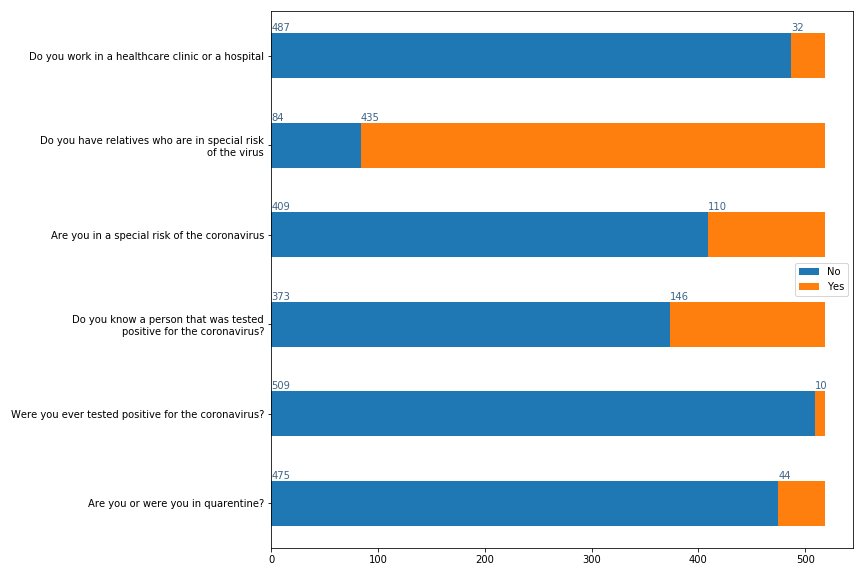}
  \caption{The distribution of the binary variables in our survey. }
  \label{fig:binary}
\end{figure}

\end{document}